\documentstyle[12pt]{article}

\input{epsf.tex}

\title{Doing numerical cosmology with the Cactus code}

\author{Dumitru N. Vulcanov\thanks{
Permanent address : The West University of Timi\c soara,
Theoretical and Computational Physics Department,
B-dul V. P\^ arvan no. 4, 1900 Timi\c soara,  Rom\^ ania, e-mail : 
{\tt vulcan@physics.uvt.ro}}\\
Max-Planck-Institut f\" ur Gravitationsphysik\\
Albert-Einstein-Institut\\
Numerical Relativity Group\\
Golm, Am M\" uhlenberg 1, D-14476, Germany}

\begin{document}

\date{}
\maketitle

\begin{abstract}
  The article presents some aspects concerning the construction of a 
new thorn for the Cactus code, a complete 3-dimensional machinery for 
numerical relativity. This thorn is completely dedicated to numerical
simulations in cosmology, that means it can provide evolutions of 
different cosmological models, mainly based on Friedman-Robertson-Walker
metric. Some numerical results are presented, testing the convergence,
stability and the  applicability of the code.
\end{abstract}

\section{Introduction}

The core of modern cosmology is the theory of General Relativity,
(\cite{1}-\cite{2}), a 4-dimensional theory involving one
dimension of time and three of space, having as field equations the
Einstein equations :
\begin{equation}\label{EE}
R_{ij} - \frac{1}{2}g_{ij} R + \lambda g_{ij} = \frac{8 \pi G}
{c^4} T_{ij}
\end{equation}
where $\lambda$ is the cosmological constant, $R_{ij}$ the Ricci tensor, 
$R$ the Ricci scalar, $g_{ij}$ the space-time metric, $T_{ij}$ the 
stress-energy tensor, G the gravitational constant, $c$ the speed of light and
 $i,j=0,1,2,3$. Numerical Relativity (NR) is concerned with the study of 
numerical solutions of the Einstein's equations for the gravitational field 
(\cite{4}), and when we apply these equations to cosmological models we have
numerical cosmology as a branch of NR. Einstein equations are an
extremely complicated system of coupled, non-linear, partial
differential equations and solving them numerically makes enormous demands on
the processing power and memory of a computer. The Cactus code (\cite{5}) 
has been mainly designed as a computational toolkit (freely available for the
scientific community) for simulating different systems of partial
differential equations, as Einstein equations are   The code can be 
used to simulate fully 3-dimensional systems with strong gravitational fields:
collapsing gravitational waves, colliding black-holes, neutron stars,
and other violent astrophysical processes generating gravitational waves. 

As a branch of NR, numerical cosmology can use the Cactus code for his
purposes. Although Einstein field equations reduce to a small number
of equations in most cosmological models (\cite{7}-\cite{8}), 
trying to simulate more
complex models (as models with cosmological constant or  having matter
fields coupled with gravity) it becomes difficult to compose special
numerical codes for every specific case. Thus our purpose was to develop
a piece of code to be used inside a  general numerical machinery
for solving Einstein equations, as the Cactus code is.
This article is a report on how we extended the Cactus code  for numerical
cosmology. In this purpose we developed a new thorn of the Cactus
code (``thorns'' are called the computer code packages embedded in 
Cactus code for some specific application). The starting point of this
new thorn, called ``Cosmo'' was an existent thorn of the Cactus code
(the ``Exact'' thorn) previously reported in (\cite{6}) for testing the
Cactus code on exact solutions of the Einstein equations. 

We investigated some cosmological models being exact solutions of
the Einstein equations in our previous article (\cite{6}).
Here the scale factor of the universe ($R(t)$) and thus the components
of the metric tensor $g_{ij}$ are known analytic function of time, given in
the code at the initial time as initial data. But in our new Cosmo thorn,
the main point is that we have to use the general Robertson-Walker
metric, having the scale factor of the universe ($R(t)$) as a unknown function
to be evolved during the time evolution of the Einstein equations through
the Cactus code. Thus the only job to be done by the Cosmo thorn is
to provide good initial data (that means at the initial time $t_0$ considered
the actual time of the universe, for example) and then let the Cactus
code to solve the Einstein equations through the time. In addition
it is necessary to prescribe convenient boundary values for the 
functions and variables involved. This is necessary because the Cactus 
code has a limited boundary methods implemented, not appropriate for our  
numerical cosmology purposes.

The article is organized as follows : the next section  2, presents some 
general facts about the way we built the Cosmo thorn, how we solved the 
problems exposed above and the necessary theoretical cosmology background. 
Section 3 is dedicated to a matter dominated universe model  based on 
Robertson-Walker metric (\cite{7}), with pressure-less matter - ``dust'' 
and without cosmological constant. Models with cosmological constant are 
presented in the next section 4 where we concentrated again only on  
models having matter as a pressure-less dust. A wide larger class of 
possible initial data is possible  including the  so called ``coasting 
cosmologies'', models not having as a starting point a Big Bang. In all 
cases studied  and reported in this article we investigated the convergence 
and the stability of the Cactus code, in the same way we presented in 
detail in \cite{6}.

As a major conclusion we have to point out the good second order convergent 
and stable behavior of the code, even in the cases we studied both forward 
and backward time evolutions. Future perspective are open for developing 
this thorn for inflationary models. It is in our view to study those 
models having one or more scalar field coupled with gravity to control
the behavior of the model and to simulate an accelerated late time
evolution, actually called ``cosmic acceleration'' being well proved
experimentally in the last years. Thus together with other similar
thorns being in preparation at AEI, we shall have a new class of applications
for the Cactus code, a major instrument for new studies in numerical
cosmology and in cosmology, in general.

Being included in the list of thorns of the Cactus code, actually the files
for our thorn are available by request, directly at the author's e-mail
address and in the near future it will be included in the CVS repository
of the Cactus code (see \cite{5}) for free download.

The simulations were performed
mainly on single processor machines, using both the AEI computer
network (SGI or Dec machines with UNIX operating system) and a Pentium
III machine with a UNIX FreeBSD operating system at the West University of 
Timi\c soara.  Some of the simulations, with similar results, were also done
on the Origin 2000 supercomputer at the AEI. 

Through this article and in the  Cactus code we shall use geometrical units 
with $G=c=1$.

\section{The Cactus code,
the Robertson-Walker metric and the Cosmo Thorn}

As we mentioned earlier, in modern cosmology we are using the Robertson-Walker
metric (RW), namely  (\cite{7}-\cite{8})
\begin{equation}\label{RW}
ds^2 = -c^2 dt^2 + R(t)^2 \left [ \frac{dr^2}{1-k r^2} + r^2 (d\theta^2 +
\sin^2 \theta~d\phi^2)\right ] 
\end{equation}
or in isotropic coordinates :
\begin{equation}\label{RWI}
ds^2 = -c^2 dt^2 + \frac{R(t)^2}{(1+\frac{k}{4} r^{\prime 2})^2}  \left [ 
dr^{\prime 2} + r^{\prime 2} (d\theta^2 +
\sin^2 \theta~d\phi^2)\right ] 
\end{equation}
where
\begin{eqnarray}
r^{\prime} = \frac{2 r}{1+\sqrt{1-k r^2}}\nonumber
\end{eqnarray}
as a generic metric for describing the dynamics of the universe through
the Einstein equations (\ref{EE}). Here $k$ is a constant with arbitrary
value, positive (for closed universes), negative (for open universe) and
zero for flat universes. Usually, this constant is taken $1$, $-1$ or
$0$ respectively. When we consider all the matter in the universe as a 
perfect fluid, the stress-energy tensor  may be given as
\begin{equation}\label{TMUNU}
T_{ij}=(\epsilon+p)u_{i}u_{j}+p g_{ij} 
\end{equation}
and is necessary to prescribe a state equation, i.e. a relation
between the pressure $p$ and the density of the universe $\rho$. For
matter dominated universes usually the pressure is $p=0$ and this will
be the case we concentrated in our simulations.

As our main purpose was to build a new thorn for the Cactus code, generically 
named from now one ``The Cosmo thorn'' for evolving the RW metric above
in all three cases for $k=-1, 0$ and $+1$. The main problem was to 
to "convince" Cactus to evolve, without having the time behavior of $R(t)$ 
function (actually, $R(t)$ is a function of the thorn). Although we 
used the same strategy as in the above mentioned Exact thorn (see \cite{6})
here, not having an analytic function of time for $R(t)$ the solution
was to provide Cactus code  all initial data he needs, namely $R(t)$ 
{\bf and $\dot{R}(t)= dR(t)/dt$} at the initial time $t_0$ we choose ! 
As a result, new routines were added for calculating the extrinsic
curvature \cite{3} components at the initial time, namely :
\begin{equation}\label{extri}
K_{\alpha \beta} = - 2 R(t)^2\frac{\dot{R}(t)}{R(t)}\delta_{\alpha \beta} = 
-2 R(t)^2 H(t) \delta_{\alpha \beta}
\end{equation}
where $H(t)$ is the {\bf Hubble constant} function (!). Note that we have
here as lapse function, $N=1$ and a shift vector as $N_{\alpha}=0$, 
$\alpha,\beta = 1,2,3$ as always in RW 
cosmology.

Another item we concentrated on was the boundary problem : - Cactus code
and the thorns which solve numerically the Einstein equations (namely
ADM\_BSSN and ADM - see \cite{5}) have not implemented proper boundary 
conditions for the RW metric. In our  previous article \cite{6} on  Exact
thorn we used ``flat'' boundary condition but this is not appropiate here !
To solve this problem we used a simple hint : $R(t)$ and 
$H(t)$ are functions of time only, thus they are constant on the numerical
grid at one time, having the same value inside the grid and on the boundary.
As a result, we added a new boundary method  through a new value of the
parameter {\bf bound} of the ADM\_BSSN thorn (see \cite{5} and \cite{6}),
called {\bf external} which activates some new routines we composed in
this purpose, having as a generic name {\bf recover}. In these routines
the spatial metric components $g_{\alpha \beta}$ and the extrinsic curvature 
components $K_{\alpha \beta}$ on the boundaries are  ``recovered'' from the 
values of the $R(t)$ and $H(t)$ on the interior points by calculating them 
using equations (\ref{RW}) and (\ref{extri}). 
Also these new  boundary conditions are necessary during the intermediate
steps of integration used by ADM\_BSSN thorn in the so-called 
``IterativeCN.F'' routine (where a three steps Cranck-Nicholson integration 
method is applied). For this purpose we slightly modified the 
``IterativeCN.F'' routine for ``injecting'' our specific ``recover'' 
boundary values. Actually this was possible through a facility of the Cactus
code which makes possible to call different routines from different thorns
using special ``include'' commands (see \cite{5}).

\section{Friedman-Robertson-Walker (FRW) cosmologies}

By ``FRW cosmologies'' we mean those cosmological models based on
the above RW metric and being solutions of the Einstein equations (\ref{EE})
without cosmological constant ($\lambda =0$) - see for example (\cite{7})
and (\cite{8}). Restricting ourselves
only to matter dominated universes (i.e. $p(t)=0$) as we mentioned earlier,
and solving the conservation law of the perfect fluid which emulates
the matter, we have for the stress-energy tensor components
\begin{equation}\label{Stress}
T_{\alpha \beta} = 0 \hbox{~~and~~} T_{00} = \frac{c^2 C}{R(t)^3}
\end{equation}
where $C$ is a constant depending on the the mass density of the universe at 
the initial time $t_0$. Solving the Einstein equations  for this 
constant and at the initial time $t_0$ where $R(t_0)=R_0$ and $H(t_0) = H_0$, 
we have 
\begin{equation}\label{constant}
C= \Omega_0 \rho_c R_0^3 = 2 q_0 \rho_c R_0^3 = q_0 c^2 \frac{3 H_0^2 R_0^3}
{4 \pi G R(t)^3}
\end{equation}
where $ q_0$ is the deceleration parameter at the initial time, i.e.
\begin{equation}\label{decel}
q(t)=-\frac{\ddot{R}(t)}{H(t)^2 R(t)}
\end{equation}
and $\Omega_0$ is the so-called density factor, defined as
\begin{equation}\label{densfact}
\Omega_0 = \frac{\rho_0}{\rho_c}
\end{equation}
$\rho_0$ being the initial density of the universe and $\rho_c$ the critical
density (i.e. the density of a flat three-dimensional universe where
$k=0$). It is proven (\cite{8}) that we have :
\begin{equation}\label{relations}
\Omega_0 = 2 q_0 \hbox{~~~and~~~} \rho_c = \frac{3 H_0^2}{8 \pi G}
\end{equation}
In order to provide initial data for the Cactus code we need to solve
the Einstein equations at the initial time for the initial Hubble constant
$H_0$ (to calculate the values of the extrinsic curvature through
equation (\ref{extri})) and for the initial scale factor $R_0$. It is
known that we have here three solutions, depending on the geometric
factor $k=0,1,-1$. Thus, for $k=0$ (flat three-dimensional spacetime) where
\begin{eqnarray}
q_0 = \frac{1}{2} \hbox{~~~and ~~~} \Omega_0 =1 \nonumber
\end{eqnarray}
we have (see \ref{8} eq. 4.37)
\begin{equation}\label{k=0}
H_0 = \frac{2}{3 t_0}
\end{equation}
and we shall take $R_0=1$, as a value for calculating, for now one
the relative scale factor, i.e. $R(t)/R_0$ which will be our
numerical output. For the case of closed universe, $k=1$ we have 
\begin{eqnarray}
q_0 > \frac{1}{2} \hbox{~~~ and ~~~} \Omega_0 > 1\nonumber
\end{eqnarray}
In this case (as in the next one for $k = -1$) we do not
have an analytic solution for the scale factor, but for the initial
time $t_0$ we can easily obtain that (\ref{8} eq. 4.51) :
\begin{equation}\label{H_0fork=1}
H_0 = \frac{q_0}{(2q_0-1)^{3/2}}\left [ cos^{-1}\left (\frac{1-q_0}
{q_0}\right )-\frac{\sqrt{2q_0-1}}{q_0}\right ]\cdot \frac{1}{t_0}
\end{equation}
and the initial scale factor is :
\begin{equation}\label{R_0fork=1}
R_0 = \frac{c}{H_0\sqrt{2q_0-1}}
\end{equation}
Finally, for the open universes ($k=-1$) where we have
\begin{eqnarray}
q_0 < \frac{1}{2} \hbox{ ~~~and ~~~} \Omega_0 < 1 \nonumber
\end{eqnarray}
As it is shown (\cite{8} eq. 4.64), here we have :
\begin{equation}\label{H_0fork=-1}
H_0 = \frac{q_0}{(1-2q_0)^{3/2}}\left [ \frac{\sqrt{1-2q_0}}{q_0}-
ln \left (\frac{1-q_0 + \sqrt{1-2q_0}}{q_0}\right )\right ]\cdot\frac{1}{t_0}
\end{equation}
and the initial scale factor is :
\begin{equation}\label{R_0fork=-1}
R_0 = \frac{c}{H_0\sqrt{1-2q_0}}
\end{equation}
With these results reminded, we can now proceed to explain how we produced our
simulations. Taking as initial data the initial time ($t_0$) in billion
years, being an internal parameter of the Cactus code, namely
{\bf CCTK\_intitial\_time}, we then have all the variables (in our
geometrical units with $c=G=1$, of course) rescaled in billion years. 
We also have  the initial value of the deceleration parameter, $q_0$
denoted in the code as {\bf robson\_q}. As  output we have the 
scale factor, more precisely the relative scale factor
$R(t)/R_0$ (denoted in the code as {\bf raza} ) and the Hubble parameter
function H(t) (denoted in the code as {\bf hubble}. For evolving these
two functions we produced two internal routines of the thorn, which
mainly calculates their values from the general output values provided
by the Cactus code and the ADM\_BSSN thorn (as the extrinsic curvature and
the three-dimensional metric components).

Here we shall present the 
results we obtained on evolving a matter dominated universe (i.e 
pressure-less matter, $p=0$) based on the RW metric. Some explanations
are necessary in order to clarity our figures ( see also (\cite{6})).  
Here and in the next figures, ``normalized'' means the L2 norm of the 
respective function calculated using all the values of the function on 
the computational grid at one time and the respective output Cactus files 
are denoted with {\bf \_nm2} extension.

\begin{figure}
\epsfxsize=3.0in
\epsfysize=3.0in
\epsfbox{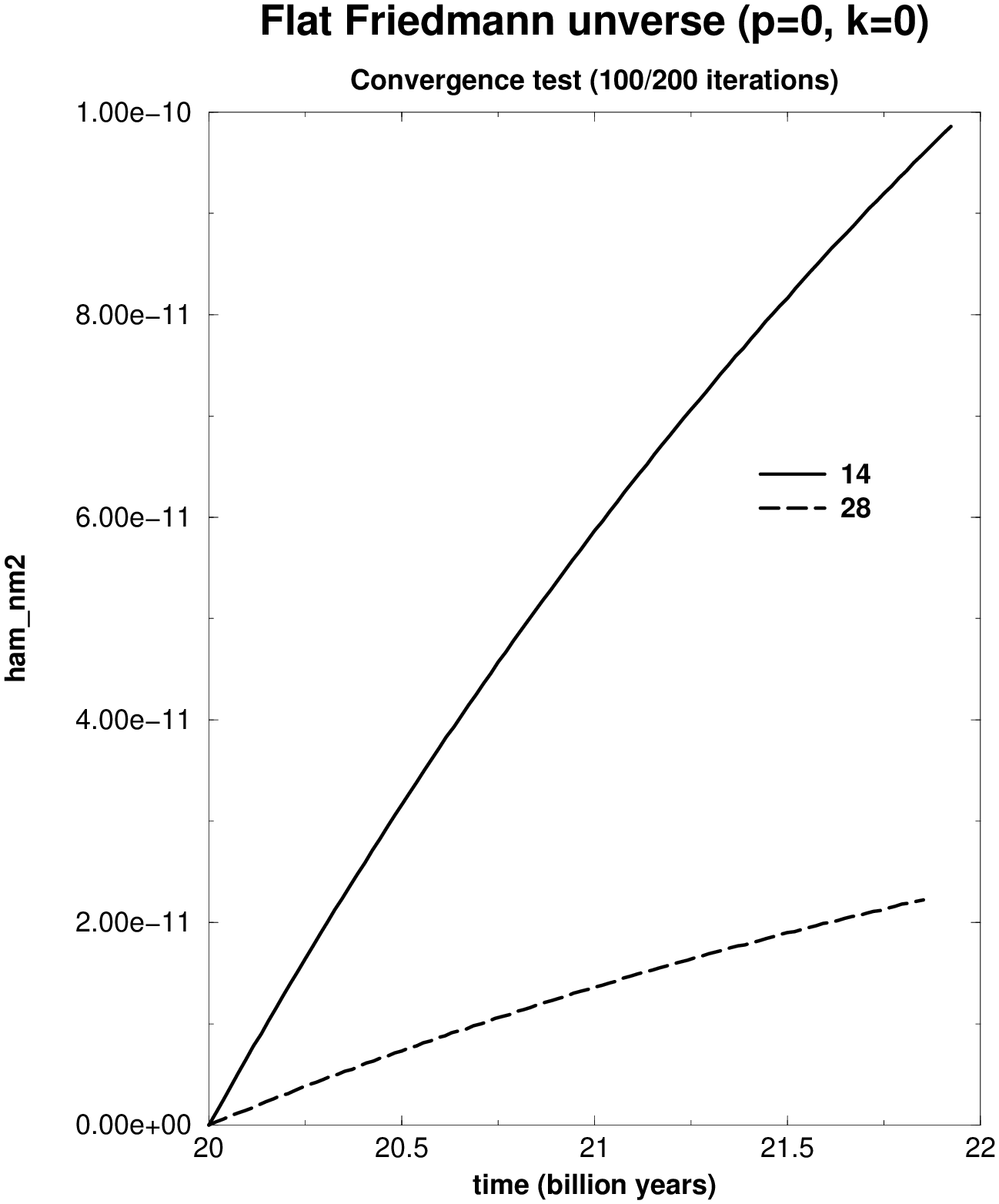}
\vspace{-3.0in}
\hspace{3.2in}
\epsfxsize=3.0in
\epsfysize=3.0in
\epsfbox{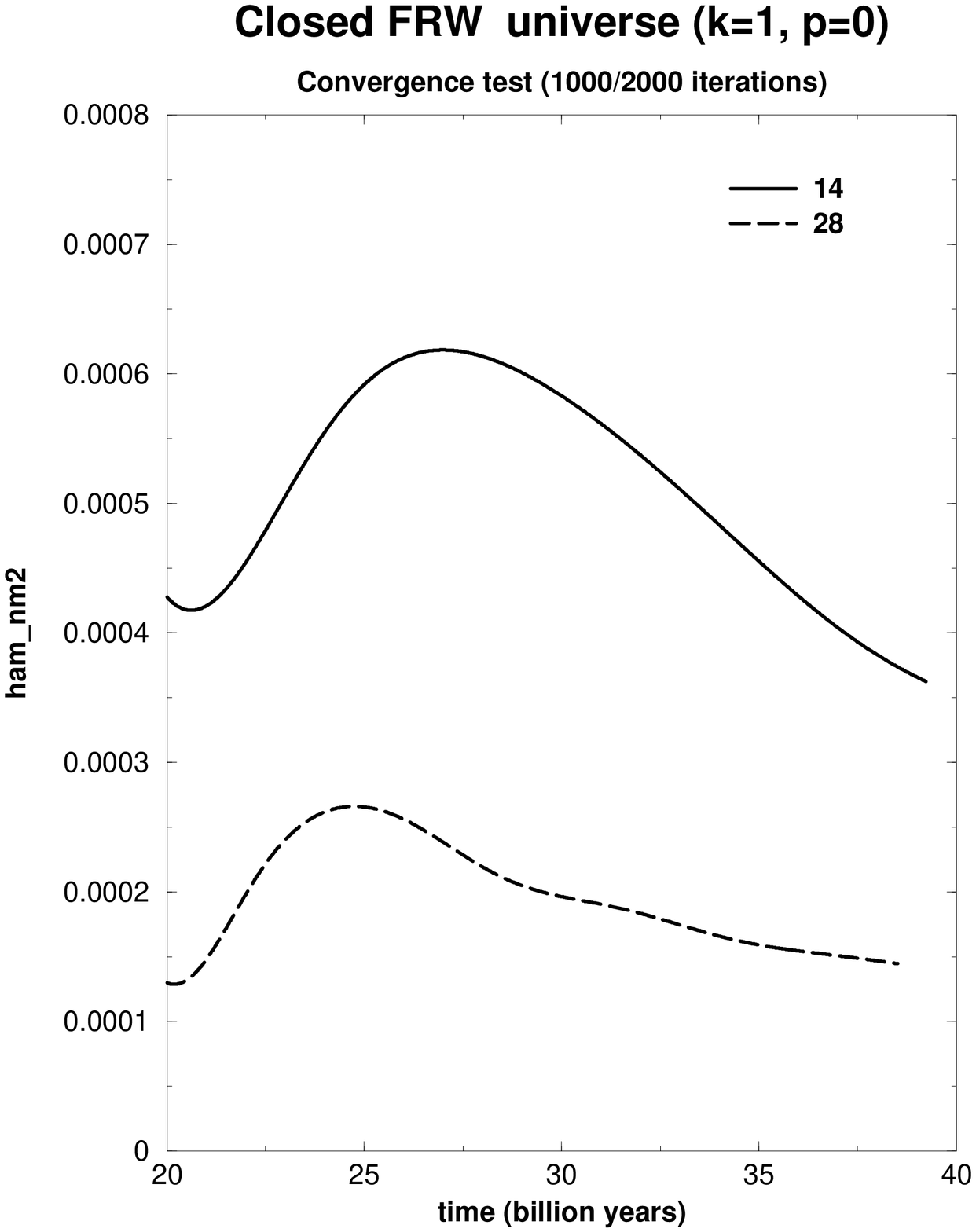}
\caption{Convergence test for FRW flat matter dominated universe (k=0, p=0) for
100/200 iterations (left side) and for FRW closed matter dominated universe (k=1, p=0) 1000/2000 iterations (right side)}
\label{fig:fig1}
\end{figure}

Cactus code is evolving the Einstein equations using the 3+1 decomposition
of space-time (both in ADM or BSSN evolution methods -see (\cite{3}) and 
(\cite{10})). Thus the Einstein
equations can be split in two groups : the dynamic equations (for the
time derivatives of the three-dimensional metric and extrinsic curvature)
and the constraint equations (the Hamiltonian constraint and the
Momentum constraint) - see \cite{1} and \cite{3}.  The constraint equations
 are  satisfied during all time evolution
of the system. Thus one of the main tests on the convergence in the 
Cactus code is given by the time behavior of the Hamiltonian constraint
(an output of the Cactus code through a thorn called ADMConstraints, 
namely {\bf ham}). In some of the next figures (here and in the next sections)
we show the convergence of the L2 norm of the Hamiltonian
constraint for  different number of iterations, using
two different resolutions on the grid, one with $14^3$ and the second
with $28^3$ points (both grids cover the same region of spacetime, so
the grid with more points has a smaller value of the finite difference
interval $\Delta x$).  Notice that the Hamiltonian constraint for a
true solution of the Einstein equations  should be equal to zero.  
Finite differencing errors imply that the numerical solution will have 
a non-vanishing value of the Hamiltonian constraint.  For a consistent finite
difference approximation of the Einstein equations, 
we should expect the Hamiltonian
constraint to approach zero as the resolution is increased.  For a
second order approximation, the value of the Hamiltonian constraint
should go down by a factor of four when the resolution is doubled.
This was the method to test the convergence of the code in all examples
presented in this article. We have obtained good second order convergence
in all examples as is shown in the next figures. 

First of our figures (\ref{fig:fig1} and ~\ref{fig:fig2}) are pointing out
the convergence of the code for different examples of FRW cosmologies. The
time is scaled in billion years (as for all figures in this section) 
starting with an initial time of 20 billion years. As it is obvious from
these figures a good second order convergence is visible, even for
long time evolutions. For the case of the closed universe ($k=1$) we
had to restrict our simulations only till around 9000 iterations (for
a grid with $14^3$  points) because in this case the universe is evolving
to a Big Crunch (opposite to the Big Bang) as it is pointed out in the
right panel of the next figure (\ref{fig:fig3}). Here the scale factor
is evolving to a maximum having the value
\begin{eqnarray}
R_{max} = \frac{2q_0}{(2q_0-1)^{3/2}} \cdot \frac{c}{H_0}\nonumber
\end{eqnarray}
twice of the actual value $R_0$ (we used here $q_0=1$). For spatially flat
universes we used, of course, $q_0=0.5$ and for open universes we had
$q_0=0.25$.

\begin{figure}
\epsfxsize=3.0in
\epsfysize=3.0in
\epsfbox{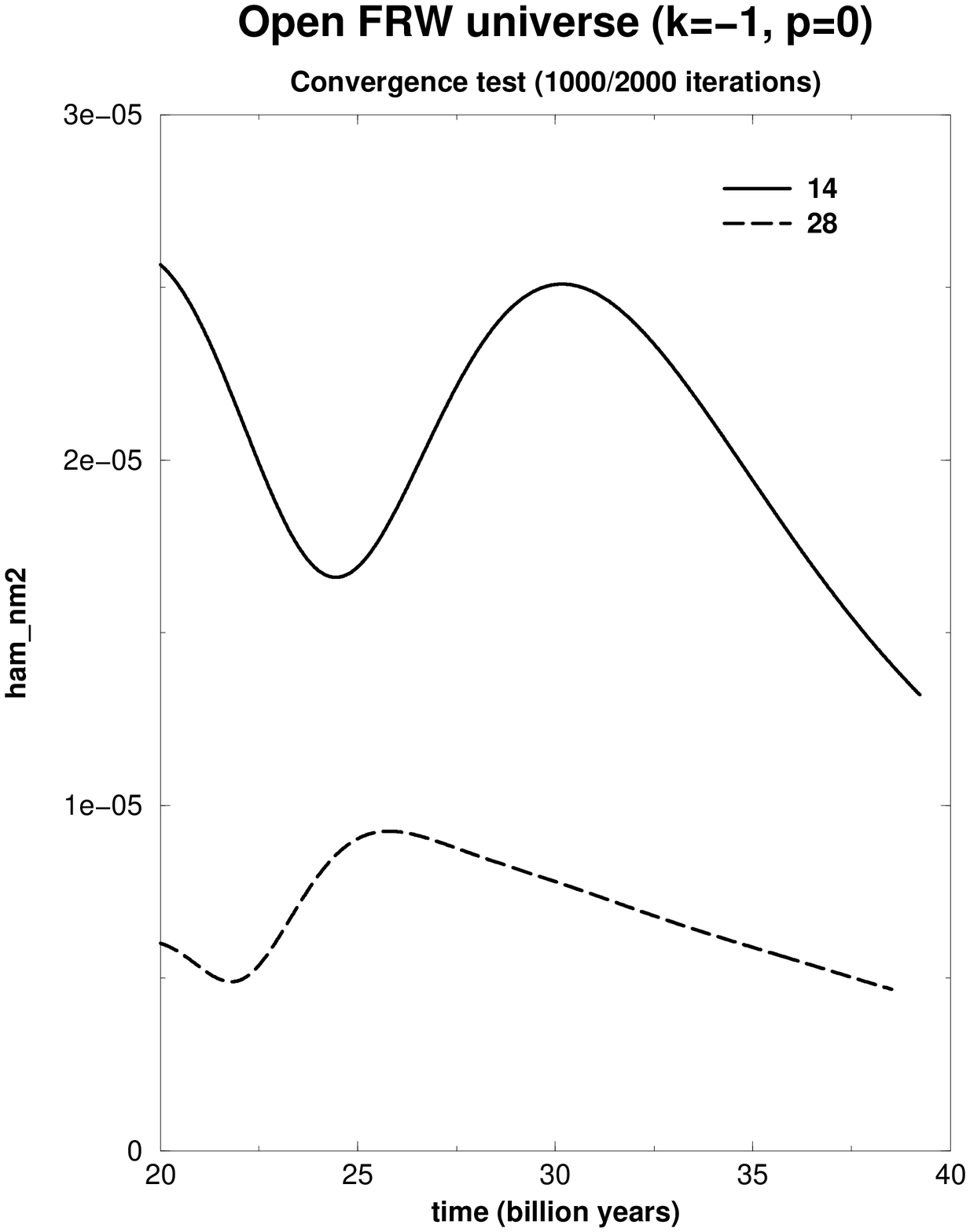}
\vspace{-3.0in}
\hspace{3.2in}
\epsfxsize=3.0in
\epsfysize=3.0in
\epsfbox{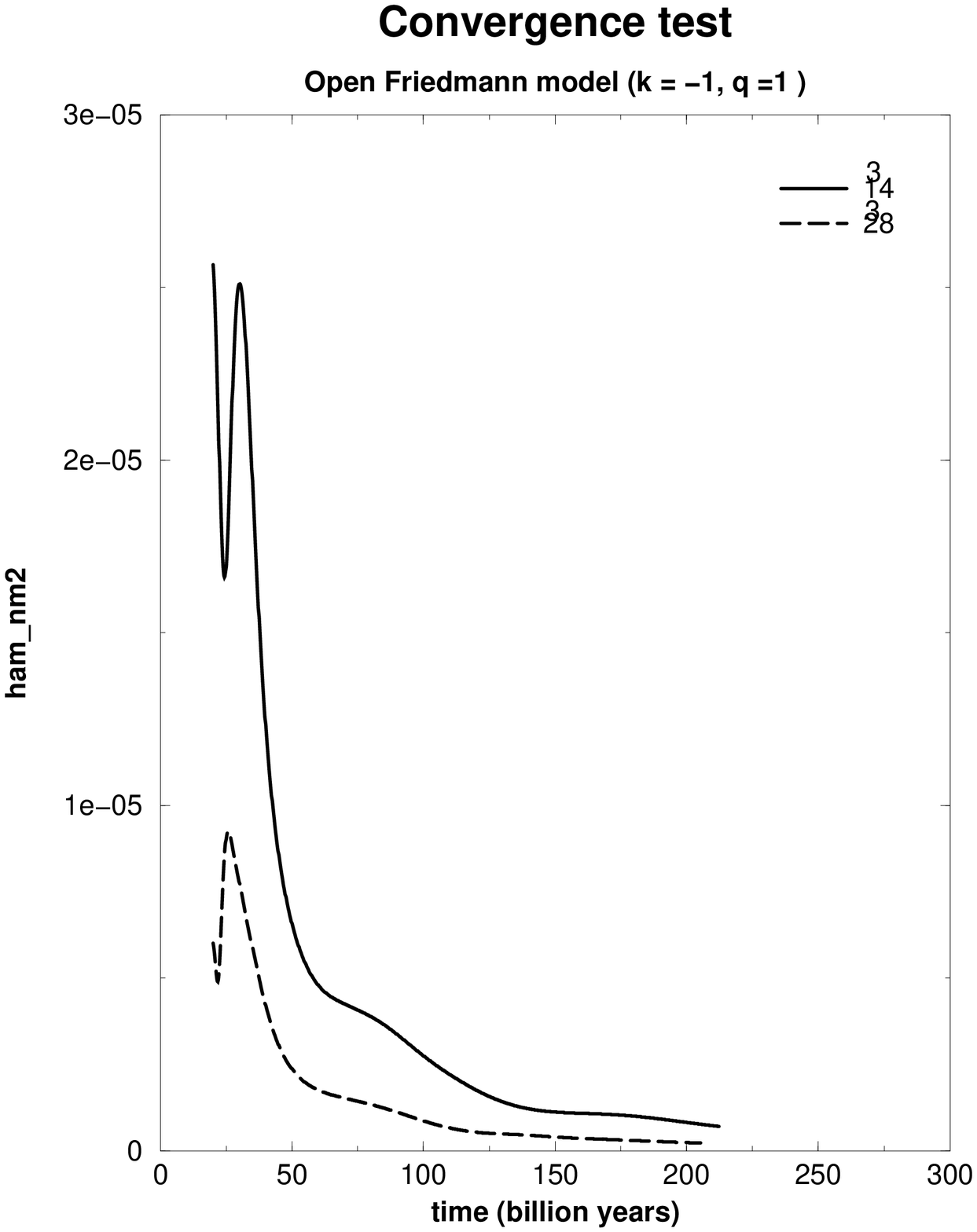}
\caption{Convergence test for FRW open matter dominate universe (k=-1, p=0) for
1000/2000 iterations (left side) and for 9000 iterations (right panel)}
\label{fig:fig2}
\end{figure}

Figure (\ref{fig:fig3}) contains also in his left panel the evolution
of the Hubble parameter function in time. We plotted here, for convenience
the modulus of this function being calculated through the L2 norm as
is output by the code. 

\begin{figure}
\epsfxsize=3.0in
\epsfysize=3.0in
\epsfbox{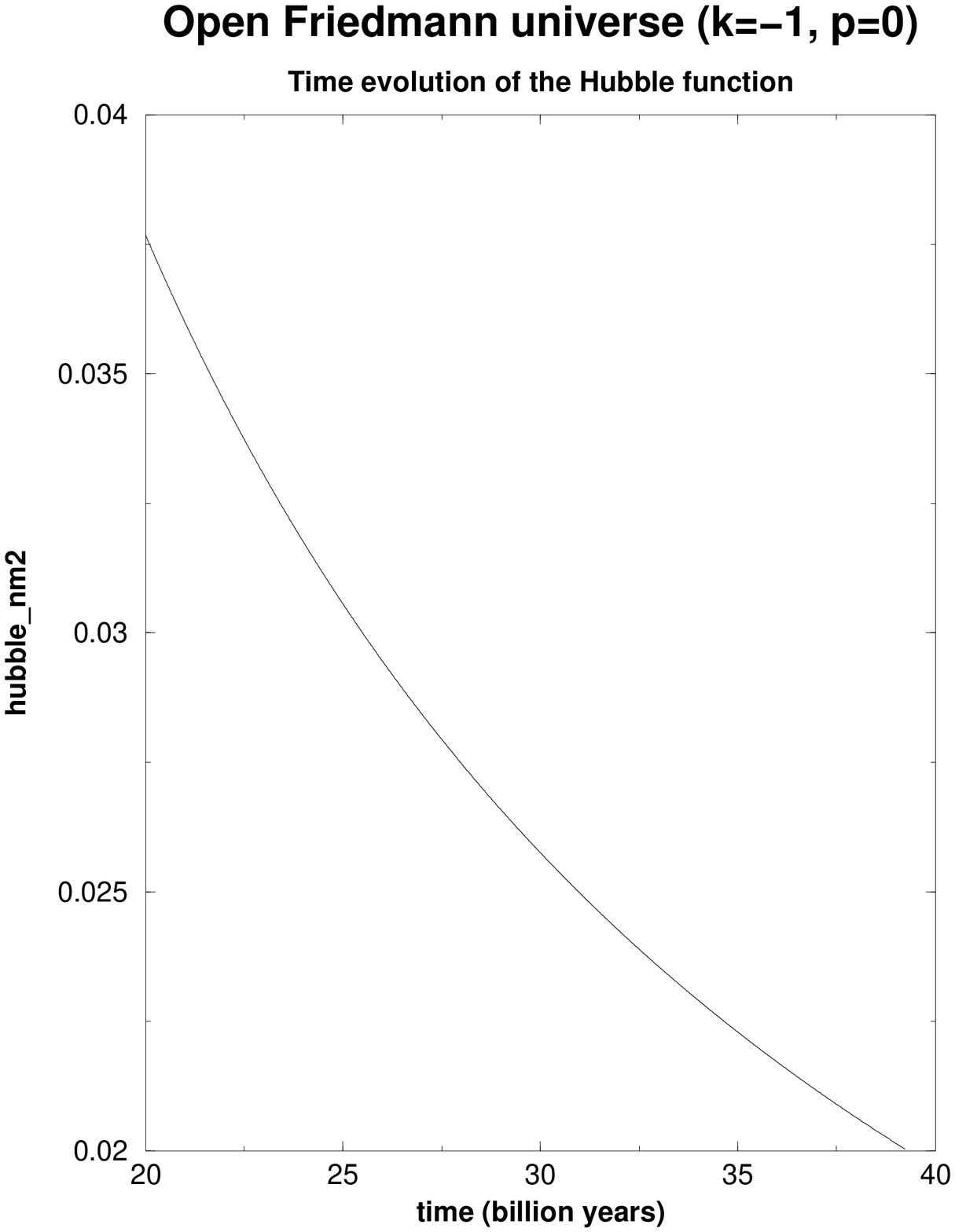}
\vspace{-3.0in}
\hspace{3.2in}
\epsfxsize=3.0in
\epsfysize=3.0in
\epsfbox{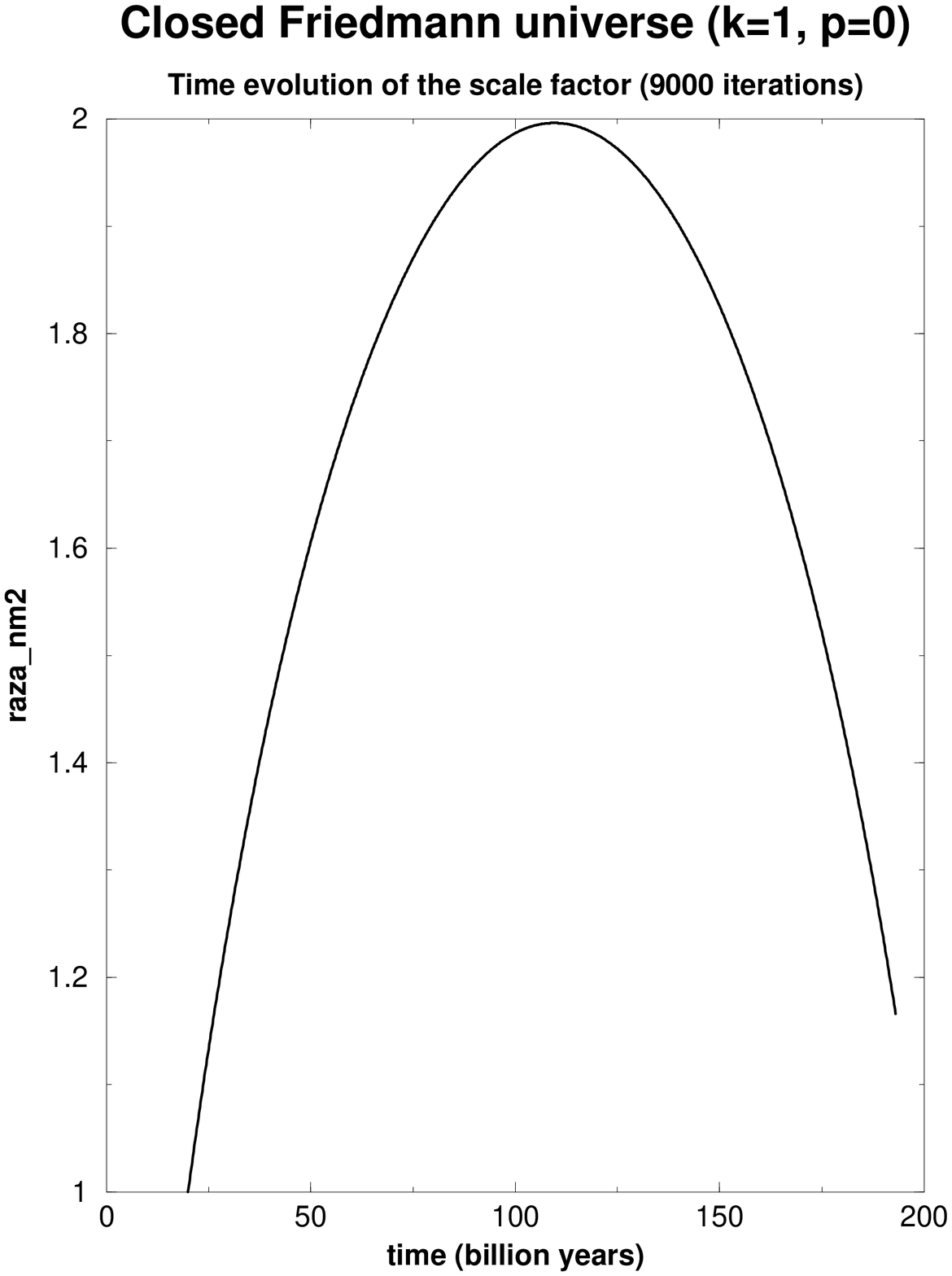}
\caption{Time evolution of the Hubble function 
for FRW open matter dominated universe (k=-1, p=0)
1000/2000 iterations (left panel) and  of the relative scale factor for
for FRW closed matter dominated universe (k=+1, p=0)
9000 iterations (right panel)}
\label{fig:fig3}
\end{figure}

\begin{figure}
\epsfxsize=3.0in
\epsfysize=3.0in
\epsfbox{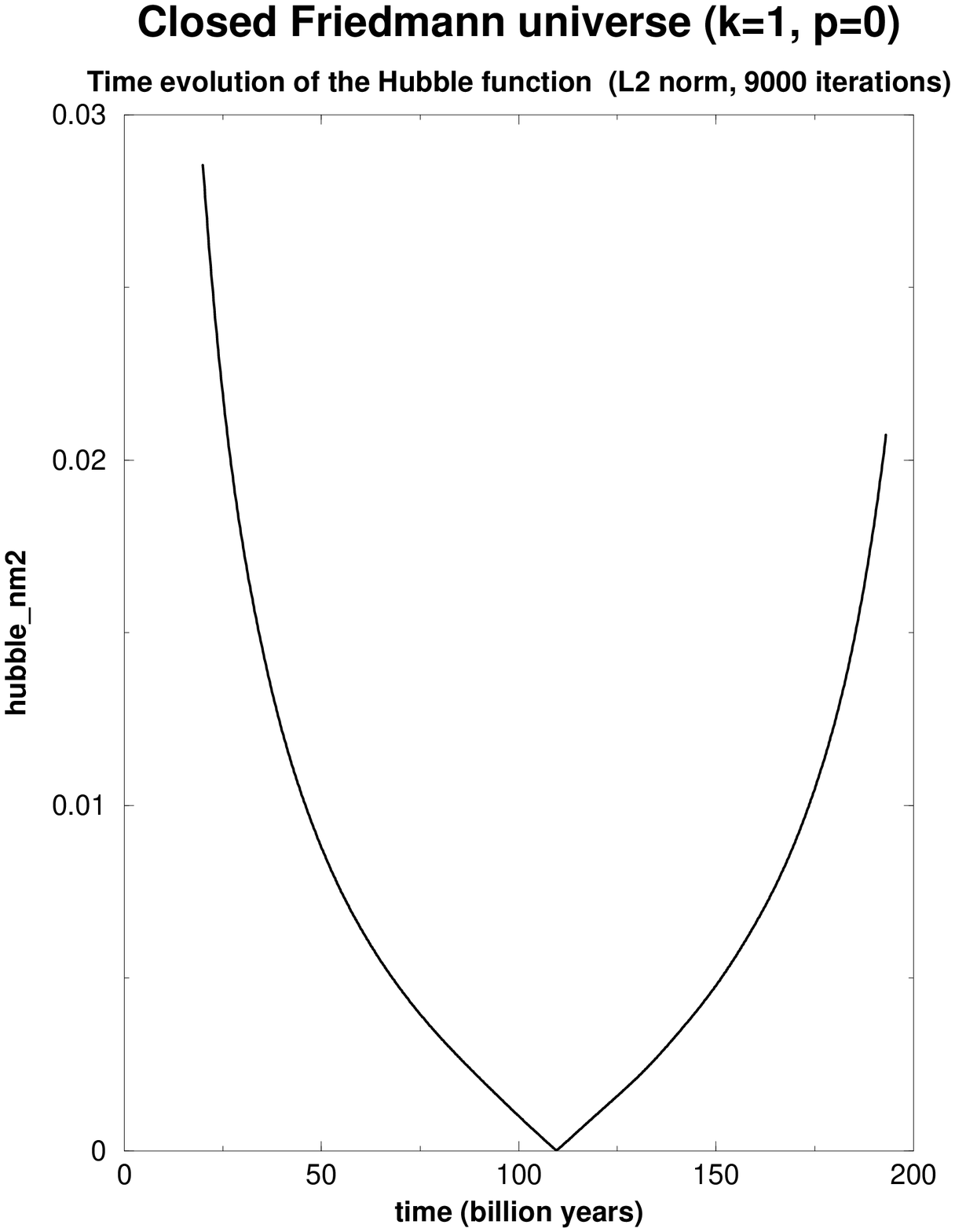}
\vspace{-3.0in}
\hspace{3.2in}
\epsfxsize=3.0in
\epsfysize=3.0in
\epsfbox{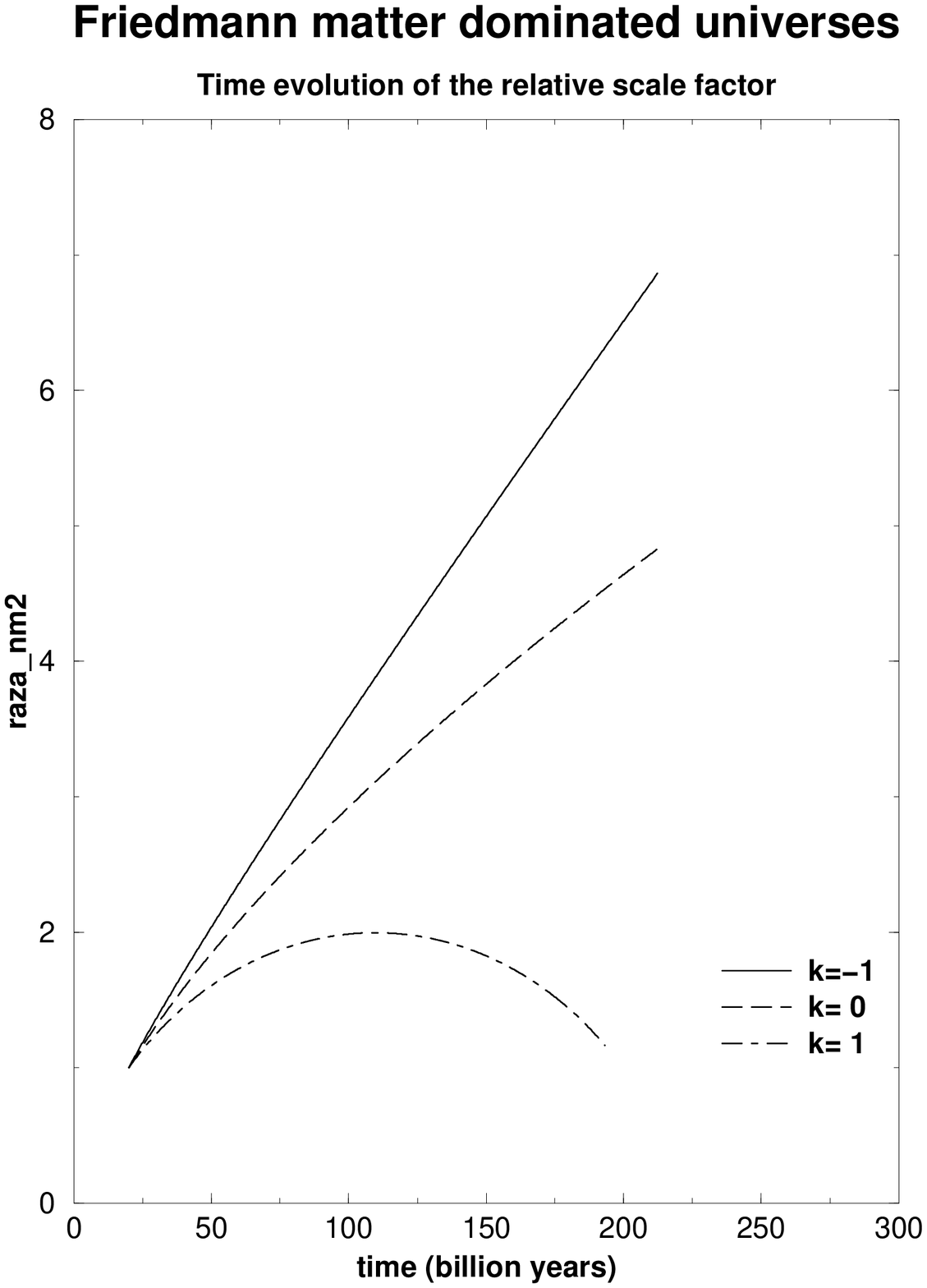}
\caption{Time evolution of the Hubble function (L2 norm)
for FRW closed matter dominated universe (k=+1, p=0)
9000 iterations (left panel) and the time evolution of the relative 
scale factor for  FRW matter dominated universes (p=0, k=+1,0,-1)
9000 iterations (right panel)}
\label{fig:fig4}
\end{figure}

The same comment is valuable for the left panel of the figure (\ref{fig:fig4})
where we can point out also the coincidence of the minimum zero value
of the Hubble parameter with the maximum of the scale factor from the
previous figure (\ref{fig:fig3}). 
The right panel of (\ref{fig:fig4}) is dedicated
to the simultaneous plot of the three cases, pointing out the differences
between the geometries of the three models.

\begin{figure}
\epsfxsize=3.0in
\epsfysize=3.0in
\epsfbox{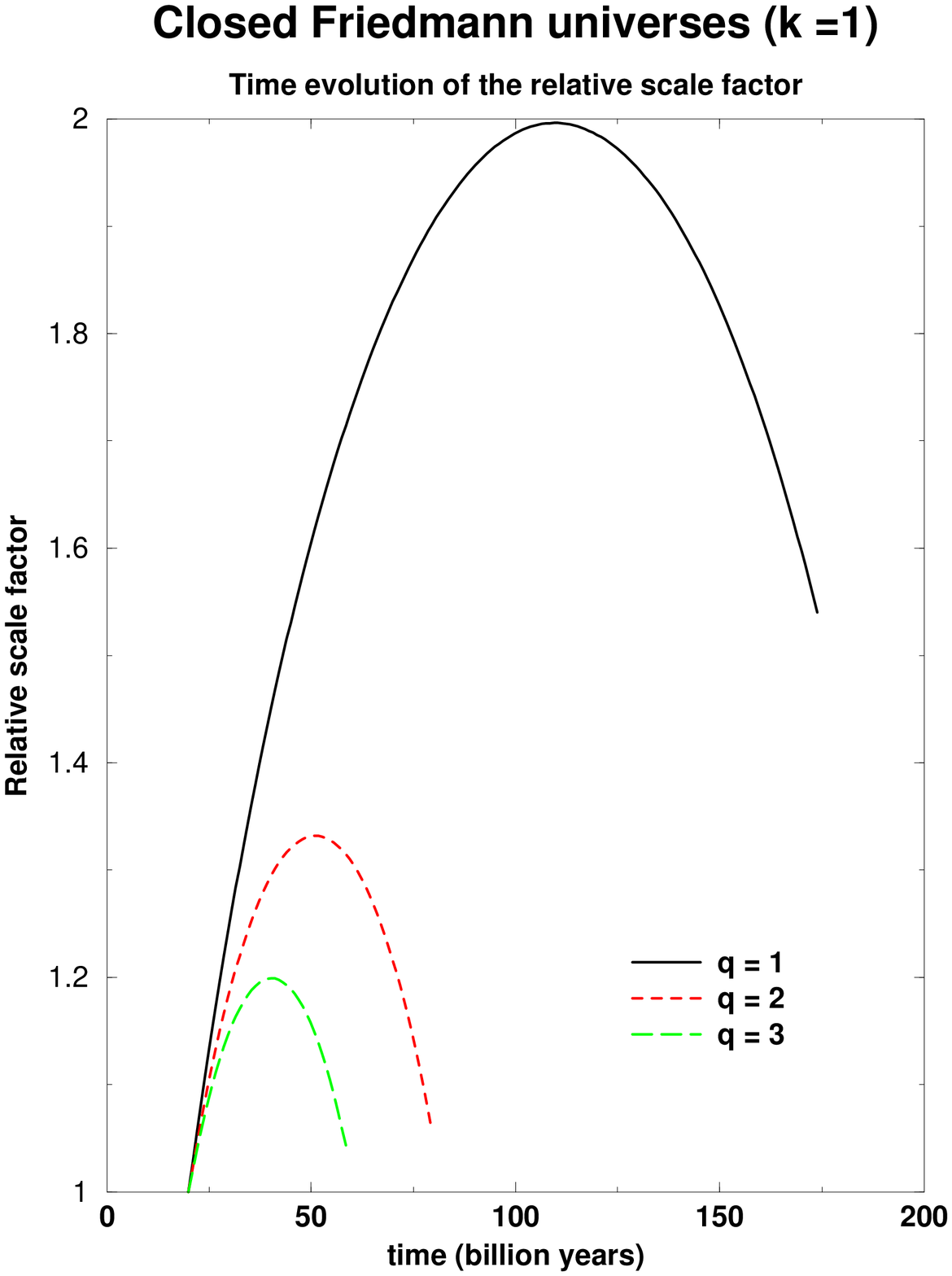}
\caption{Time evolution of the relative scale factor for
for different closed matter dominated FRW universes (k=+1, p=0)
9000 iterations}
\label{fig:fig5}
\end{figure}

Last of our figures from this section, (\ref{fig:fig5}) contains the
plots of the relative scale factor time evolution for different closed matter 
dominated FRW universes, having different deceleration parameter 
(i.e. different density parameters). It is visible here that as the 
deceleration parameter (density parameter) is increasing so is decreasing 
the time-life of the closed universe between a Big-Bang and a Big-Crunch.

\section{FRW cosmologies with cosmological constant}

Here we shall deal with cosmologies based on the above RW metric but
having a non-vanishing cosmological constant. In spite of the failure of the
the initial tentative of Einstein to use the cosmological constant
as a trick to fit general relativity with, at that time presumed, static
universe, the cosmological constant
is again used in modern cosmology in order to establish more realistic 
models of the universe, as the inflationary models and, recently, the
so called cosmic acceleration. Early attempts (see (\cite{9})) on 
using numerical 
routines for solving the Einstein equations for these models were reported.
Because our goal is to implement in the Cactus code a thorn for doing similar
simulations, but on a larger scale and fully three-dimensional
we shall refer to these previous results only for comparing our results.
Meanwhile we shall use here the same notations and scaling of the parameters
as in (\cite{9}) for prescribing the initial data sets for the Cactus code.

This time I will  proceed in a slightly different way than in the previous
section, scaling all our variables and parameters in terms of  the
initial value of the Hubble parameter function $H_0$, which is not
completely free in our universe, having been measured to within a factor
of 2. Thus for the cosmological constant $\lambda$ for 
instance, we use $\lambda/H_0^2$ as a basic parameter, denoted in 
the code with {\bf robson\_l}. Of course, for dimensional scaling reasons 
we prescribed at the initial time $H_0=1$. Thus, for example, at the 
output we  have the relative value of the Hubble function $H(t)/H_0$.
Meanwhile we need a  parameter to describe the density of the universe,
and this time we  used $\Omega_0$ instead of the deceleration parameter
(remark that here we have no more the relation $\Omega_0 = 2 q_0$). We have
also the Gaussian curvature $K(t)$ defined as
\begin{eqnarray}
K(t)= \frac{k}{R(t)^2}\nonumber
\end{eqnarray}
which will be, of course identical to the geometric factor $k$ at the
initial time  when we take the initial scale factor $R_0=1$ as we 
done in our simulations. Following the line indicated in (\cite{9}) we can 
define the density factor as
\begin{equation}\label{densfact1}
\Omega(t) = \frac{8 \pi G}{3} \frac{\rho(t)}{H(t)^2} \hbox{~~~and at the
initial time~~~} \Omega_0=\frac{8 \pi G}{3}\frac{\rho_0}{H_0^2}
\end{equation}
where $\rho_0$ is the initial density of the matter in the universe.
Now solving the Einstein equations at the initial time and after some
algebraic manipulations we obtain (for $c=G=1$, $R_0=1$) :
\begin{equation}\label{kappa}
K_0 = k = H_0^2 \left (\Omega -1 +\frac{1}{3}\frac{\lambda}{H_0^2}
\right )
\end{equation}
and for the stress-energy tensor (where we included also the term containing
the cosmological factor, as we did in (\cite{6})) :
\begin{equation}\label{stresslambda}
T_{ij}= T^0_{ij}-\frac{c^4}{8\pi G}\lambda g_{ij}
\hbox{~~~where~~~}
T^0_{00}= \frac{3}{8\pi G}\Omega_0 H_0^2 \frac{R_0^3}{R(t)^3}
\hbox{~~~;~~~}
T^0_{\alpha\beta}=0 ~~~~~\forall \alpha,\beta=1,2,3
\end{equation}
Now  our numerical results are pointed out in the next figures. First 
we show, in figure (\ref{fig:fig6}) the time evolution of the scale
factor for some cosmological models, having the density factor $\Omega$
being $0.1$ (left panel) and $1.0$ (right panel). The models are 
differentiated by their respective cosmological constant (more precisely
the factor $\lambda/H_0^2$). For the time scale we used here Hubble
time units, i.e. the time is expressed as $t/t_H$ where $t_H$ is
the Hubble time : $t_H = H_0^{-1}$. In these simulations, as in the next ones
we started from the actual time (taken here $t_0=0$) and we investigated
the time behavior of the model in both directions of time, forward, as
well as backward in time. 

\begin{figure}
\epsfxsize=3.0in
\epsfysize=3.0in
\epsfbox{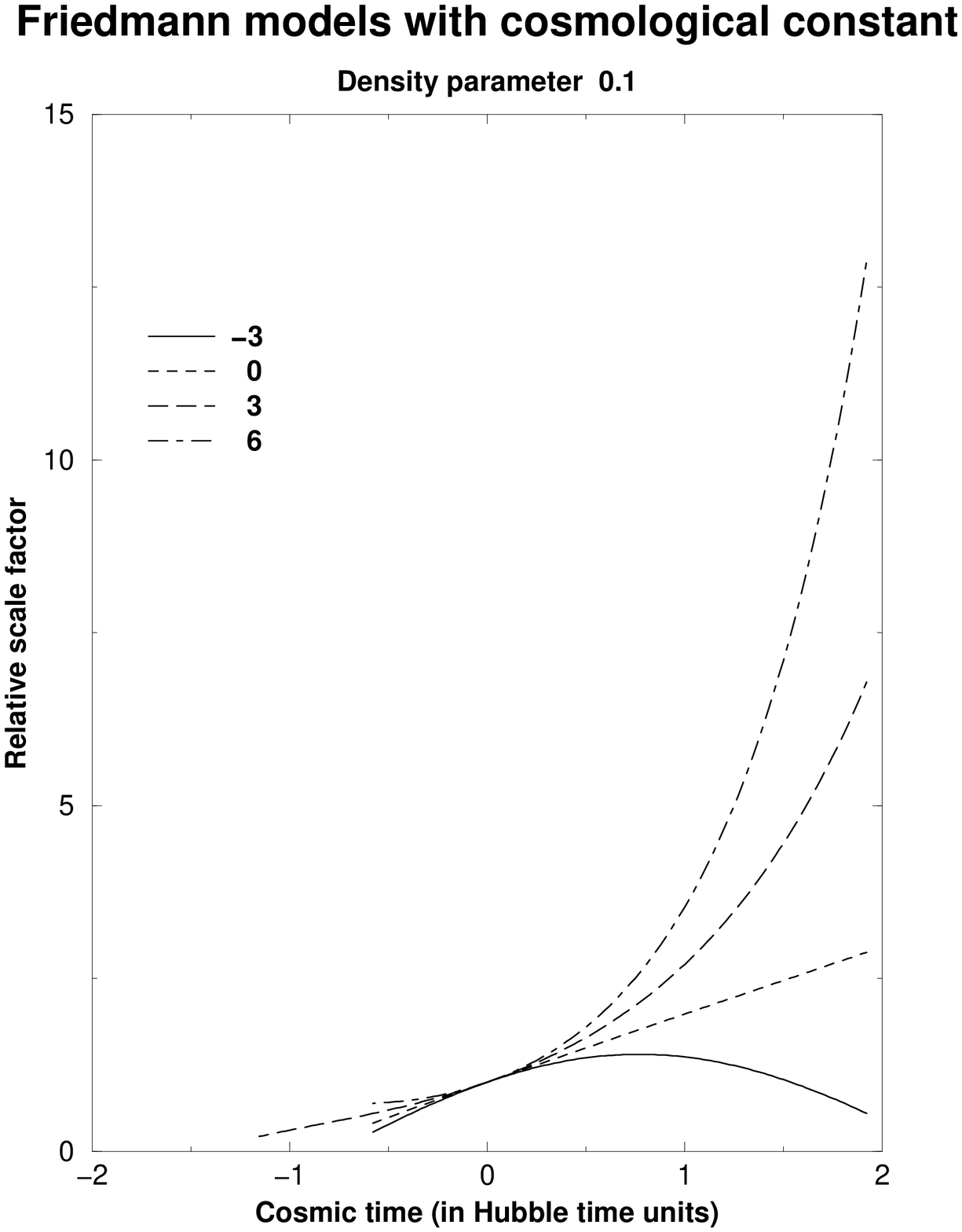}
\vspace{-3.0in}
\hspace{3.2in}
\epsfxsize=3.0in
\epsfysize=3.0in
\epsfbox{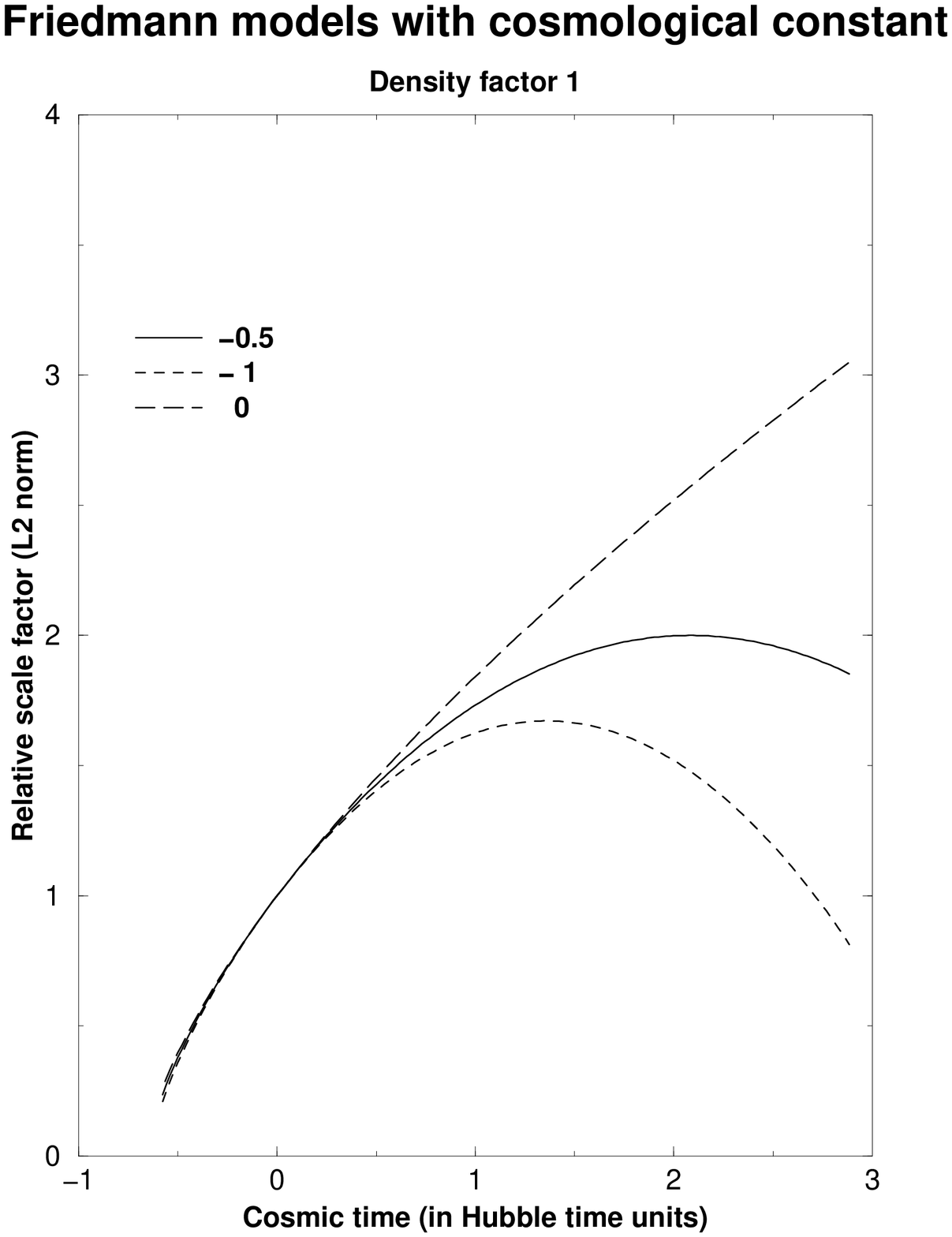}
\caption{Time evolution of the scale factor for FRW models with cosmological
constant. The left panel represent four different models for a density 
factor $\Omega=0.1$ and $\Lambda/H_0^2=\{-3,0,3,6\}$ and the right panel 
represent models for a density factor $\Omega=1.0$ and $\lambda/H_0^2
=\{-1,-0.5,9\}$}
\label{fig:fig6}
\end{figure}

Next figure (\ref{fig:fig7}) shows, in his left panel some special cases
of FRW models with cosmological constant, where the scale factor shows off 
a different evolution : namely it starts somewhere back in time with a 
certain finite value,
it decreases in time, then it has a period (longer or shorter) of constant
value, evolving finally as an expanding model (of De-Sitter type, for example).
 These models, having a period of constant value of the scale factor, are
sometimes called ``coasting cosmologies'' and are investigated as an
alternative (realistic or not) to the standard Big-Bang model, not being
generated from a Big-Bang, as can be seen  from our simulations.
The right panel of the figure (\ref{fig:fig7}) contains the time evolution
of the Hubble constant function (the L2 norm, as it is output by the
Cactus code, otherwise the values after the null minimum value being negative
ones) for a model with $\Omega=0.1$ and $\lambda/H_0^2=5$.

\begin{figure}
\epsfxsize=3.0in
\epsfysize=3.0in
\epsfbox{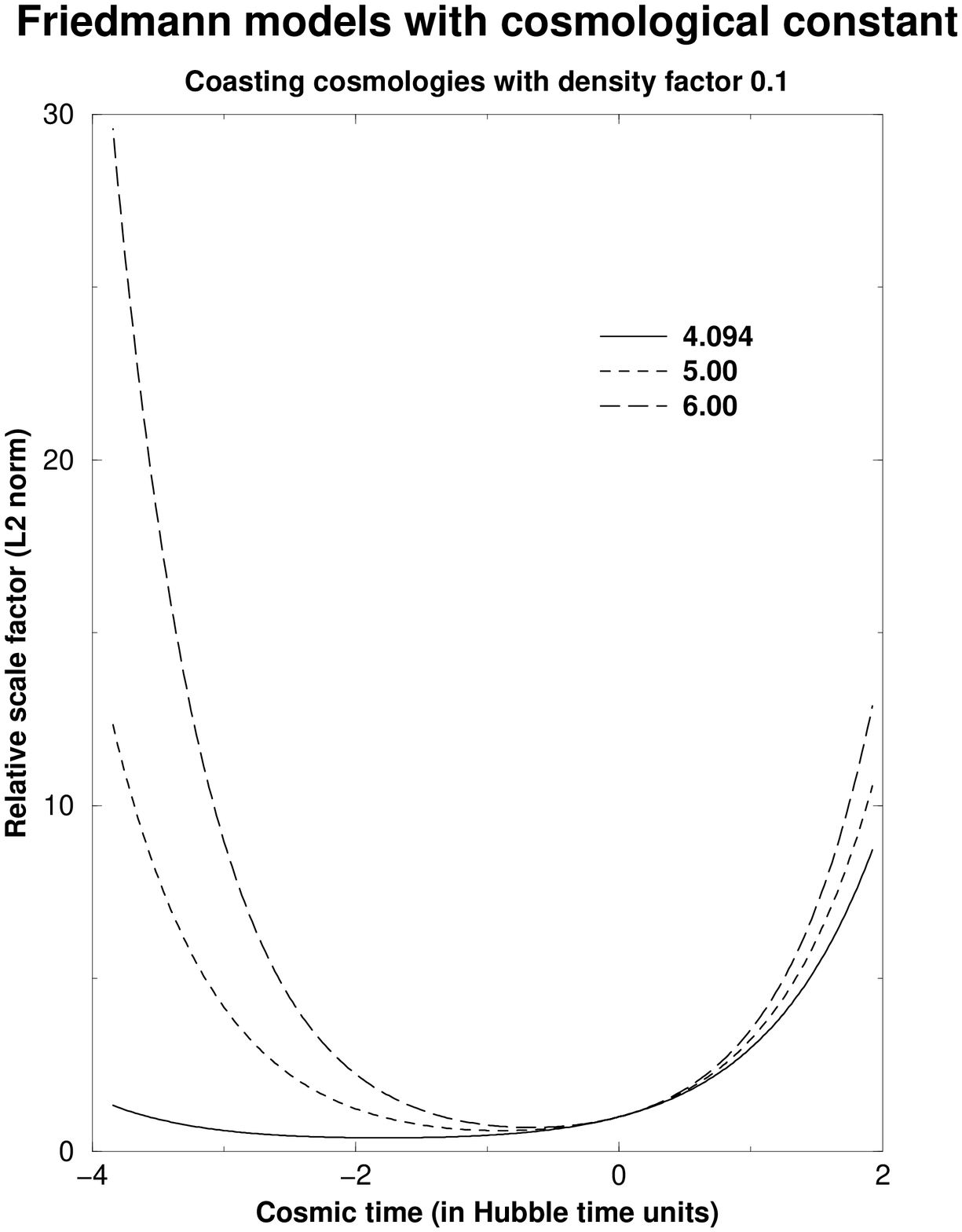}
\vspace{-3.0in}
\hspace{3.2in}
\epsfxsize=3.0in
\epsfysize=3.0in
\epsfbox{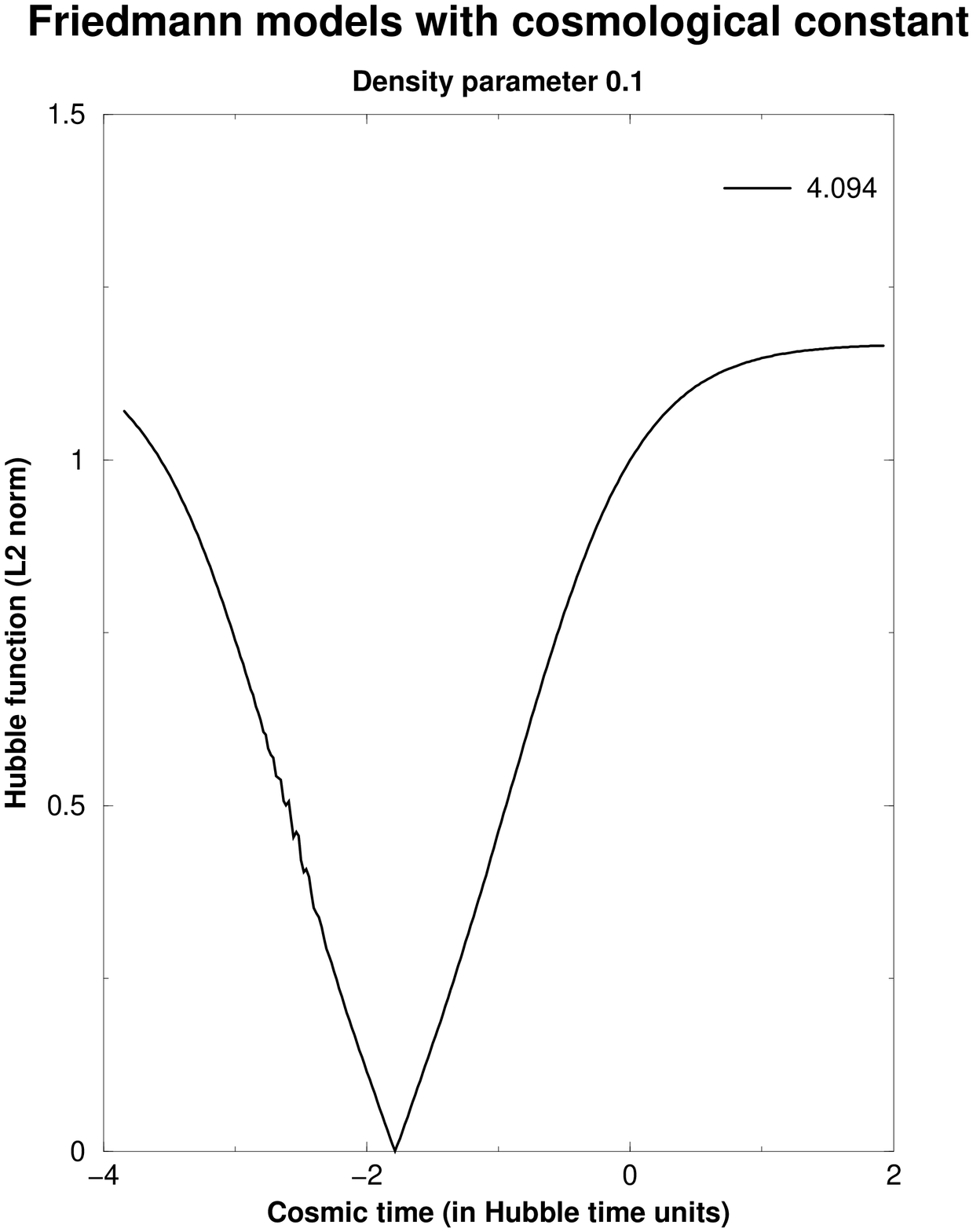}
\caption{``Coasting'' FRW cosmologies for $\Omega = 0.1$ and $\lambda/H_0^2=
4.94, 5, 6$ (left panel) and the time evolution of the Hubble constant
function for a model with $\Omega=0.1$ and $\lambda/H_0^2=5$ (right panel)}
\label{fig:fig7}
\end{figure}

We must point out that we performed here also several convergence tests, as
in the  previous section, obtaining again good second order convergence.
As an example we shall present in the last figure ({\ref{fig:fig8})
the convergence test for $14^3$ and $28^3$ points on the grid for a model
having again $\Omega=0.1$ and $\lambda/H_0^2=5$ for an evolution forward in
time (left panel) and backward in time (right panel).

\begin{figure}
\epsfxsize=3.0in
\epsfysize=3.0in
\epsfbox{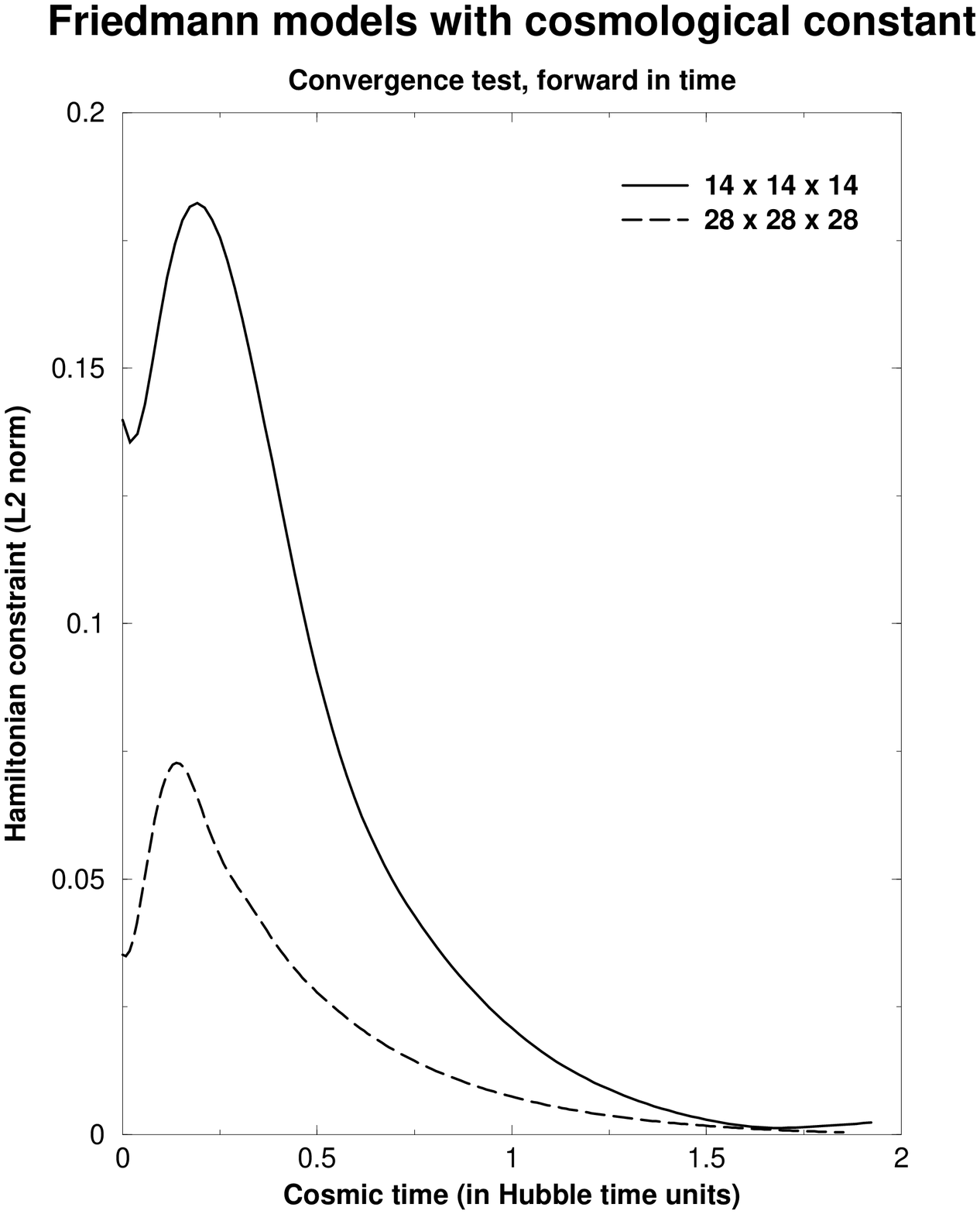}
\vspace{-3.0in}
\hspace{3.2in}
\epsfxsize=3.0in
\epsfysize=3.0in
\epsfbox{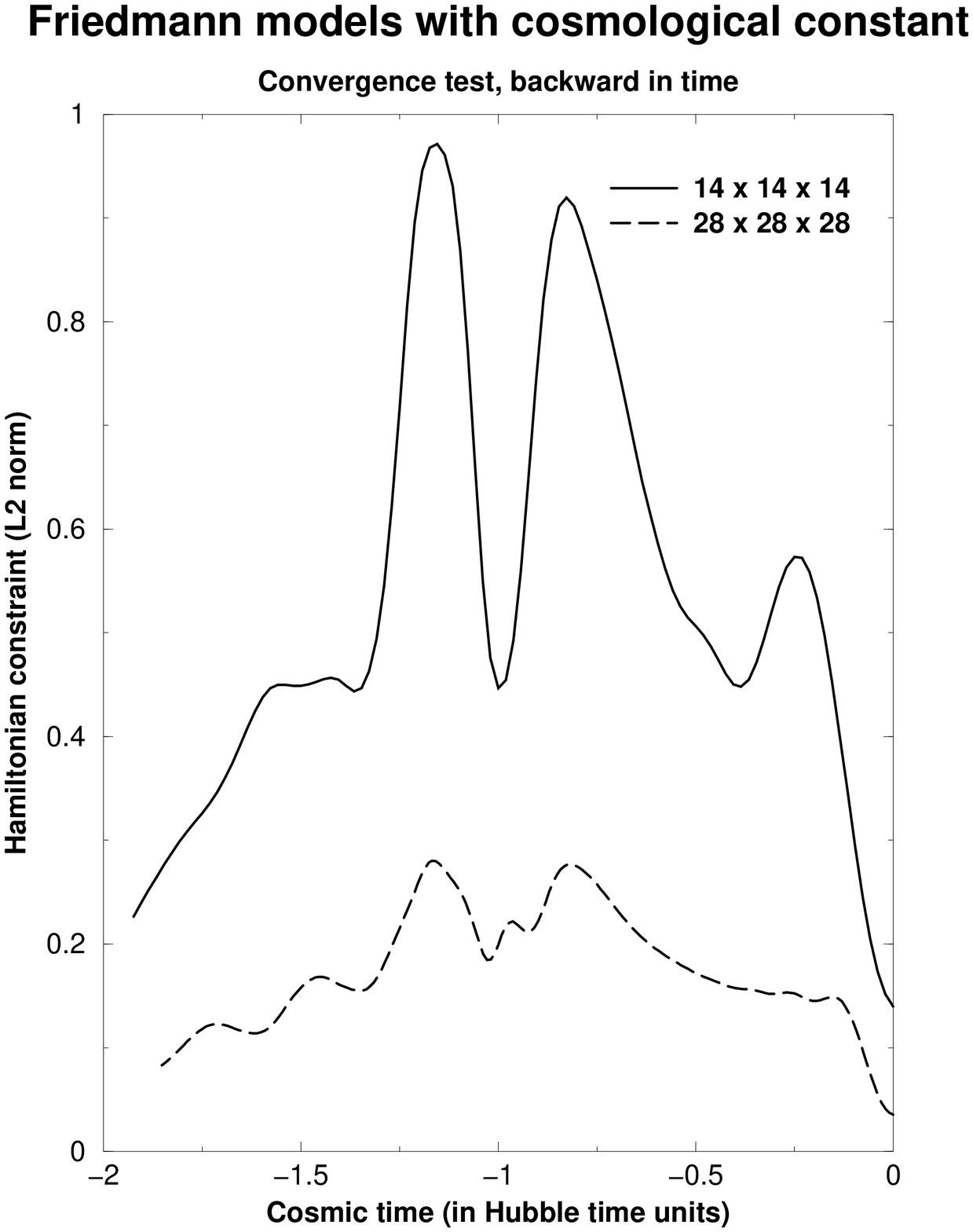}
\caption{Convergence test for $14^3$ and $28^3$ points on the grid for a model
having again $\Omega=0.1$ and $\lambda/H_0^2=5$ for an evolution forward in
time (left panel) and backward in time (right panel) and for 100/200
iterations}
\label{fig:fig8}
\end{figure}

\section*{Acknowledgments}
The author is   deeply indebted to Prof.
E.~Seidel for patience, continuous support and encouragement, and to prof. 
Schutz for his understanding and help. The friendly hospitality and 
environment the author experienced at AEI during his visits is 
kindly acknowledged.

\end{document}